\documentclass{webofc}
\usepackage[varg]{txfonts}   % Web of Conferences font
\usepackage{hyperref,multirow}
\usepackage{url}
\hypersetup{colorlinks=true,citecolor=blue,urlcolor=blue,linkcolor=blue}
\begin{document}
\title{The properties of $\phi(2170)$ and its three-body nature}
%
% subtitle is optionnal
%
%%%\subtitle{Do you have a subtitle?\\ If so, write it here}

\author{\firstname{A.} \lastname{Mart\'inez Torres}\inst{1}\fnsep\thanks{\email{amartine@if.usp.br}} \and
        \firstname{Brenda} \lastname{B. Malabarba}\inst{1}\fnsep\thanks{\email{brenda@if.usp.br}} \and
        \firstname{K. P.} \lastname{Khemchandani}\inst{2}\fnsep\thanks{\email{kanchan.khemchandani@unifesp.br}}
        % etc.
}

\institute{Universidade de Sao Paulo, Instituto de Fisica, C.P. 05389-970, Sao 
Paulo, Brazil.
\and
Universidade Federal de Sao Paulo, C.P. 01302-907, Sao Paulo, Brazil.
}

\abstract{%
We summarize the results we obtained for the partial decay widths of $\phi(2170)$ into two-body final states formed by a $\bar K$ and a Kaonic resonance, like $K(1460)$, $K_1(1270)$, as well as to final states constituted by a $\phi$ and an $\eta/\eta^\prime$ mesons. The results obtained are compared with the values extracted from experimental data on the corresponding branching ratios, which were determined by the BESIII collaboration. A reasonable agreement is found, which together with the previous reproduction of the mass, width, and cross section for the process $e^+ e^-\to \phi f_0$ strongly indicates the molecular nature of $\phi(2170)$ as a $\phi K\bar K$ system.}
\maketitle
\section{Introduction}
Since its discovery in 2006 by the BaBar collaboration, several experimental collaborations have been trying to understand the properties of the $\phi(2170)$ meson~\cite{BaBar:2006gsq,BaBar:2007ceh,BES:2007sqy,Belle:2008kuo, BESIII:2018ldc,BESIII:2020gnc,Belle:2022fhh}. Recently, the BESIII collaboration~\cite{BESIII:2020vtu,BESIII:2020gnc,BESIII:2021bjn} have determined the product between the decay width of $\phi(2170)\to e^+ e^-$ and the branching fraction of $\phi(2170)\to \bar K K_R,~\phi\eta,~\phi\eta^\prime$, with $K_R$ being a Kaonic resonance, from fits to the corresponding data. The results found seem to challenge the theoretical predictions for the partial decay widths of $\phi(2170)$ to the same $\bar K K_R,~\phi\eta,~\phi\eta^\prime$ channels obtained within a $s\bar s$, hybrid or tetraquark picture for its inner structure~\cite{BESIII:2020vtu,BESIII:2020gnc,BESIII:2021bjn,Malabarba:2020grf,Malabarba:2023zez}. In Ref.~\cite{MartinezTorres:2008gy}, the $\phi K\bar K$ system was studied considering interactions in s-wave, and the solution of the Faddeev equations was obtained for such system within the approach of Refs.~\cite{MartinezTorres:2007sr,Khemchandani:2008rk,MartinezTorres:2008kh}. As a result, the three-body $T$-matrix for the system shows the generation of a state with mass and width compatible with that of $\phi(2170)$  when the $K\bar K$ system is forming $f_0(980)$. The $e^+e^-\to \phi f_0(980)$ cross section determined by the BaBar collaboration was also well reproduced with the model of Ref.~\cite{MartinezTorres:2008gy} by implementing the final state interaction in the $e^+e^-\to\phi f_0(980)$ cross section calculated with the approach of Ref.~\cite{Napsuciale:2007wp}, which explained the background of the process, but not the signal observed for $\phi(2170)$.

Given the recent data obtained by the BESIII collaboration about some partial decay widths of $\phi(2170)$, it would be interesting to know the corresponding values determined with the model of Ref.~\cite{MartinezTorres:2008gy} and check if they agree, or not, with the experimental data. Such compatibility with the data is by no means trivial, since models considering $\phi(2170)$ as a $s\bar s$, hybrid, tetraquark, etc., do not seem to give compatible results for all the known experimental data for $\phi(2170)$, which include the previous mentioned partial decay widths, cross sections obtained from $e^+e^-$ collisions, mass, and width. For instance, the BESIII collaboration has found that the decay mode of $\phi(2170)$ to $K^*(892)\bar K^*(892)$ is suppressed as compared to other $\bar K K_R$ final states. This fact alone does not seem to be understood considering $\phi(2170)$ to be a $s\bar s$ or hybrid state.

\section{Formalism}
The partial decay widths of $\phi(2170)$ to the aforementioned channels not only depend on the nature of $\phi(2170)$, but also on that of the Kaonic resonances present in the final state, like $\mathbb{K}\equiv K(1460)$, $K_1(1270)$, $K_1(1400)$. Having a good description of the properties of these latter states is relevant to having reliable partial decay widths for $\phi(2170)$. In the case of $K(1460)$, we consider the model of Ref.~\cite{MartinezTorres:2011gjk} in which the state is described from the $KK\bar K$ interaction, with a large coupling to the $Kf_0(980)$ configuration. In the case of $K_1(1270)$ and $K_1(1400)$ we consider three different approaches: (1) In Ref.~\cite{Geng:2006yb}, the $K\rho$ and pseudoscalar-vector coupled channel dynamics were studied and generation of $K_1(1270)$ was found as a consequence of the superposition of two poles, one at $z_1=M-i\Gamma/2=1195-i123$ MeV and other at $z_2=1284-i 73$ MeV. In this case, no signal for $K_1(1400)$ was obtained. We call this model $A$; (2) In Ref.~\cite{Palomar:2003rb}, a tensor formalism for the vector mesons was used and $K_1(1270)$ and $K_1(1400)$ were described as states obtained from the mixing of the $K_{1A}$ and $K_{1B}$ states belonging to the nonet of axial resonances. Mixing angles of $29^\circ-62^\circ$ were shown to be compatible with the experimental data available for these states. We call this model $B$; (3) Instead of relying on the results found within the previous two models for the coupling constants of the $K_1$ states to pseudoscalar-vector meson channels, we could directly use the data on the radiative decay of $K_1(1270)$ and $K_1(1400)$ available on the particle data book to estimate such couplings. We call this model $C$.

\begin{figure}
\centering
\begin{tabular}{cc}
\includegraphics[width=0.45\textwidth]{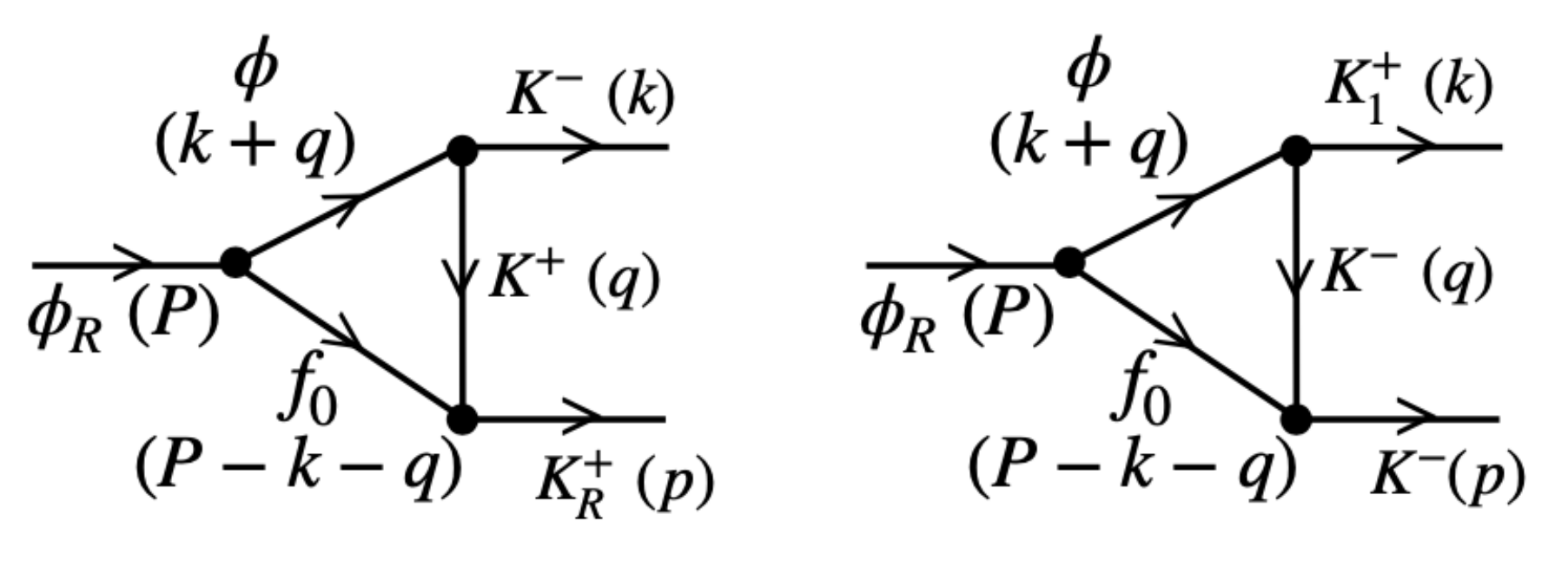}&\includegraphics[width=0.45\textwidth]{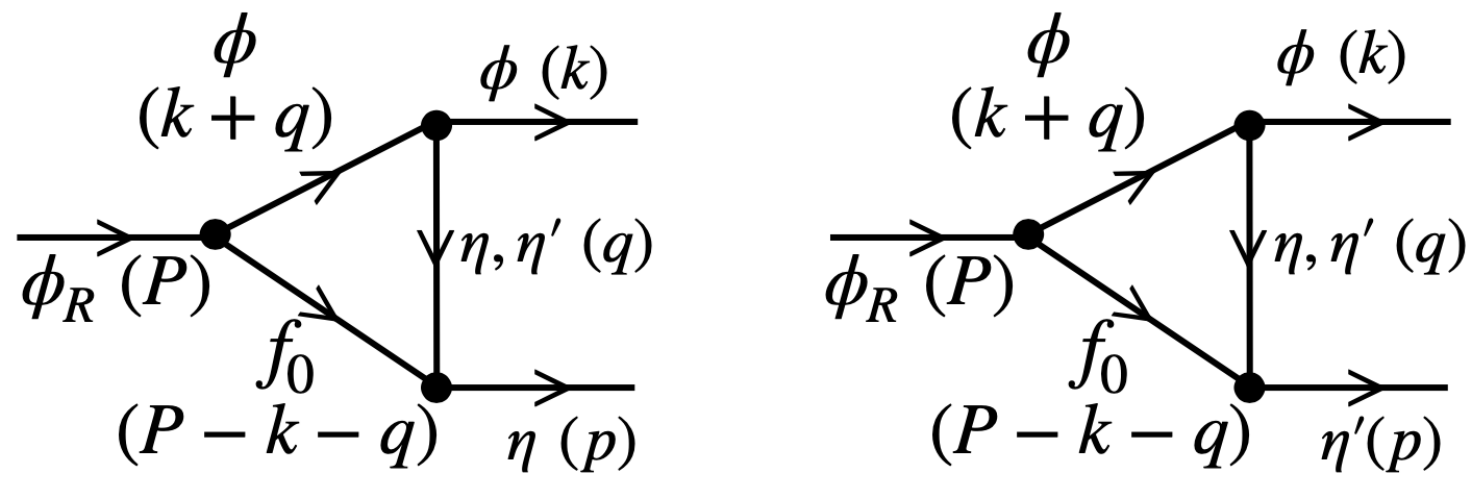}
\end{tabular}
\caption{Decay mechanism for $\phi(2170)$ to $\bar K K_R$, $K_R\equiv K(1460)$, $\bar K K_1$, $K_1\equiv K_1(1270)$, $K_1(1400)$, and to $\phi\eta$, $\phi\eta^\prime$. }\label{decay}
\end{figure}

Having in ming the coupling of $K_1(1270)$ and $K_1(1400)$ to pseudoscalar-vector channels, the molecular nature of $\phi(2170)$ as a $\phi f_0(980)$ state and that of $f_0(980)$ as a state obtained from the $K\bar K$ and pseudoscalar-pseudoscalar coupled channel dynamics~\cite{Oller:1997ti,Oller:1998hw}, the decay of $\phi(2170)\to \bar K K_R$, $\phi\eta$ and $\phi\eta^\prime$ proceeds as depicted in Fig.~\ref{decay}.

Following Refs.~\cite{MartinezTorres:2008gy,MartinezTorres:2011gjk,Oller:1997ti,Geng:2006yb}, the states $\phi(2170)$, $K(1460)$, $K_1(1270)$, and $f_0(980)$ are generated from the s-wave interactions of three or two hadron systems. Thus, the contribution of the vertices $\phi(2170)\to \phi f_0(980)$, $f_0 K^+\to K^+(1460)$, $\phi\to K^+_1 K^-$ present in the decay mechanisms of Fig.~\ref{decay} can be written as,
\begin{align}
t_{\phi_R}&=g_{\phi_R\to\phi f_0}\epsilon_{\phi_R}\cdot \epsilon_\phi,~~~t_{K_R}=g_{K^+_R\to K^+ f_0},\nonumber\\
t_{f_0\to \mathcal{P}\mathcal{P}^\prime}&=g_{f_0\to \mathcal{P}\mathcal{P}^\prime},~~~t_{K^+_1\to\phi K^+}=g_{K^+_1\to \phi K}\epsilon_{K^+_1}\cdot\epsilon_\phi,\label{ts}
\end{align}
where $g_{i\to j}$ is the coupling for the process $i\to j$, $\mathcal{P}$ and $\mathcal{P}^\prime$ represent pseudocalar particles and $\epsilon_k$ is the corresponding polarization vector for particle $k$. To determine the amplitude for the $\phi\to \mathcal{P}_1\mathcal{P}_2$ vertex, we consider the Lagrangian~\cite{Bando:1985rf}
\begin{align}
\mathcal{L}_{VPP}=-ig \langle V^\mu[\mathbb{P},\partial_\mu \mathbb{P}]\rangle,
\end{align}
with $V^\mu$ and $\mathbb{P}$ being matrices having as elements the vector and pseudoscalar meson octet fields, respectively, $g=M_V/(2f_\pi)$, $M_V\simeq M_\rho$, $f_\pi\simeq 93$ MeV, and $\langle\quad\rangle$ indicating the SU(3) trace. The coupling constant $g_{f_0\to \mathcal{P}\mathcal{P}^\prime}$ is obtained from the residue of the two-body $t$-matrix describing the interaction between two pseudoscalars. This $t$-matrix is obtained by solving the Bethe-Salpeter equation with a kernel $V$ which is determined from the lowest-order chiral Lagrangian $\mathcal{L}_{\mathbb{P}\mathbb{P}}$, implementing the $\eta-\eta^\prime$ mixing~\cite{Herrera-Siklody:1996tqr,Kaiser:2000gs,Liang:2014tia},
\begin{align}
\mathcal{L}_{\mathbb{P}\mathbb{P}}=\frac{1}{12 f^2}\langle (\partial_\mu \mathbb{P}\mathbb{P}-\mathbb{P}\partial_\mu\mathbb{P})^2+M\mathbb{P}^4\rangle.\label{LPP}
\end{align}
with
\begin{align}
\mathbb{P}&=\left(\begin{array}{ccc}A(\beta)\eta+B(\beta)\eta^\prime+\frac{\pi^0}{\sqrt{2}}&\pi^+&K^+\\\pi^-&A(\beta)\eta+B(\beta)\eta^\prime-\frac{\pi^0}{\sqrt{2}}&K^0\\K^-&\bar K^0&C(\beta)\eta+D(\beta)\eta^\prime\end{array}\right),\label{Pmat}
\end{align}
where
\begin{align}
A(\beta)&=-\frac{\text{sin}\beta}{\sqrt{3}}+\frac{\text{cos}\beta}{\sqrt{6}},~~B(\beta)=\frac{\text{sin}\beta}{\sqrt{6}}+\frac{\text{cos}\beta}{\sqrt{3}},\nonumber\\
C(\beta)&=-\frac{\text{sin}\beta}{\sqrt{3}}-\sqrt{\frac{2}{3}}\text{cos}\beta,~~D(\beta)=-\sqrt{\frac{2}{3}}\text{sin}\beta+\frac{\text{cos}\beta}{\sqrt{3}},
\end{align}
with the mixing angle $\beta$ being between $-15^\circ$ to $-22^\circ$, instead of simply considering ideal mixing (i.e., $\text{sin}\beta=-1/3$, thus $\beta\simeq -19.47^\circ$)~\cite{Lin:2021isc}, and $M$ is a matrix having as elements
\begin{align}
M&=\left(\begin{array}{ccc}m^2_\pi&0&0\\0&m^2_\pi&0\\0&0&2m^2_K-m^2_\pi\end{array}\right),
\end{align}
where $m_\pi$, $m_K$ represent the masses of the pion and the kaon, respectively. When calculating the coupling of $f_0(980)$ to $\mathcal{P}\mathcal{\bar P^\prime}$, two models were considered: (I) We use in Eq.~(\ref{LPP}) different weak decay constants for the pseudoscalars; (II) We consider a common value $f=f_\pi=93$ MeV.

We refer the reader to Refs.~\cite{Malabarba:2020grf,Malabarba:2023zez} for the values of the coupling constants involved in the vertices depicted in Fig.~\ref{decay}. Using the amplitudes of Eq.~(\ref{ts}), we can determine the contribution of the processes depicted in Fig.~\ref{decay}, which depend on different tensor integrals. As a consequence of the four-momenta dependence of the vertices, these tensor integrals can be written as $d^4q$ integrals of a numerator that depends on $q_\mu$, $q_\nu q_\mu$, etc., and a denominator which is a function of $q$, $P$ and $k$, with the latter dependence being related to the propagators of the particles in the triangular loops of Fig.~\ref{decay}. Using Lorentz covariance, we can write these tensor integrals in terms of a linear combination of $k_\mu$, $P_\mu$ or products of $k_\mu$ and $P_\mu$, depending on the order of the tensor. Such a linear combination introduces several unknown coefficients, which need to be determined.

For example, the amplitude for the process $\phi(2170)\to \phi\mathcal{P}$ depicted in Fig.~\ref{decay}, where $\mathcal{P}$ represents, in this case, an $\eta$ or $\eta^\prime$ meson,  can be written as~\cite{Malabarba:2023zez}
\begin{align}
it_{\phi_R\to\phi\mathcal{P}}&=\sum\limits_{\mathcal{P}^\prime}2g_{\phi_R\to\phi f_0}g_{f_0\to\mathcal{P}\bar{\mathcal{P}^\prime}}g_{\phi\to\phi\mathcal{P}^\prime}\epsilon^{\mu\nu\alpha\beta}\epsilon_{\phi_R\nu}(P)k_\alpha\epsilon_{\phi\beta}(k)I_\mu,\label{tImu}
\end{align}
where $I_\mu$ is the following tensor integral:
\begin{align}
I_\mu&=\int\limits_{-\infty}^{\infty}\frac{d^4q}{(2\pi)^4}\frac{q_\mu}{[(P-k-q)^2-m^2_{f_0}+i\epsilon]}\frac{1}{[(k+q)^2-m^2_\phi+i\epsilon][q^2-m^2_{\mathcal{P}^\prime}+i\epsilon]}.\label{Imudef}
\end{align}
As a consequence of the Lorentz covariance, we can write $I_\mu$ in terms of $k_\mu$ and $P_\mu$ as
\begin{align}
I_\mu=a_{\mathcal{P}^\prime} k_\mu+b_{\mathcal{P}^\prime} P_\mu,\label{Imu}
\end{align}
where $a_{\mathcal{P}^\prime}$ and $b_{\mathcal{P}^\prime}$ are the mentioned unknown coefficients. To calculate them, we proceed as follows: Multiplying Eq.~(\ref{Imu}) by $k^\mu$ and $P^\mu$, respectively, we get two coupled equations which permit to write $a_{\mathcal{P}^\prime}$ and $b_{\mathcal{P}^\prime}$ as
\begin{align}
a_{\mathcal{P}^\prime}&=\frac{P^2(k\cdot I)-(k\cdot P)(P\cdot I)}{k^2 P^2-(k\cdot P)^2},~~~b_{\mathcal{P}^\prime}=-\frac{(k\cdot P)(k\cdot I)-k^2(P\cdot I)}{k^2 P^2-(k\cdot P)^2},
\end{align}
where we have introduced 
\begin{align}
k\cdot I&=\int\limits_{-\infty}^{\infty}\frac{d^4q}{(2\pi)^4}\frac{k\cdot q}{[(P-k-q)^2-m^2_{f_0}+i\epsilon]}\frac{1}{[(k+q)^2-m^2_\phi+i\epsilon][q^2-m^2_{\mathcal{P}^\prime}+i\epsilon]},\nonumber\\
P\cdot I&=\int\limits_{-\infty}^{\infty}\frac{d^4q}{(2\pi)^4}\frac{P\cdot q}{[(P-k-q)^2-m^2_{f_0}+i\epsilon]}\frac{1}{[(k+q)^2-m^2_\phi+i\epsilon][q^2-m^2_{\mathcal{P}^\prime}+i\epsilon]}.
\end{align}
By working in the rest frame of the decaying particle, i.e., $P^\mu=(P^0,\vec{0})$, with $P^0=m_{\phi_R}$, we can express the previous integrals as
\begin{align}
k\cdot I&=\int\limits_{-\infty}^{\infty}\frac{d^3q}{(2\pi)^3}\int\limits_{-\infty}^\infty\frac{dq^0}{(2\pi)}\frac{k^0q^0-\vec{k}\cdot\vec{q}}{[(P-k-q)^2-m^2_{f_0}+i\epsilon]}\frac{1}{[(k+q)^2-m^2_\phi+i\epsilon][q^2-m^2_{\mathcal{P}^\prime}+i\epsilon]}\nonumber\\
&\equiv \int\limits_{-\infty}^{\infty}\frac{d^3q}{(2\pi)^3}[k^0\mathcal{I}_1(m_{f_0},m_\phi,m_{\mathcal{P}^\prime})-\vec{k}\cdot\vec{q}\,\mathcal{I}_0(m_{f_0},m_\phi,m_{\mathcal{P}^\prime})],\nonumber\\
P\cdot I&=P^0\int\limits_{-\infty}^{\infty}\frac{d^3q}{(2\pi)^3}\int\limits_{-\infty}^\infty\frac{dq^0}{(2\pi)} \frac{q^0}{[(P-k-q)^2-m^2_{f_0}+i\epsilon]}\frac{1}{[(k+q)^2-m^2_\phi+i\epsilon][q^2-m^2_{\mathcal{P}^\prime}+i\epsilon]}\nonumber\\
&\equiv P^0\int\limits_{-\infty}^{\infty}\frac{d^3q}{(2\pi)^3}\mathcal{I}_1(m_{f_0},m_\phi,m_{\mathcal{P}^\prime}),\label{QI}
\end{align}
where we have introduced
\begin{align}
&\mathcal{I}_n(m_1,m_2,m_3)\equiv\int\limits_{-\infty}^\infty\frac{dq^0}{(2\pi)}\frac{(q^0)^n}{[(P-k-q)^2-m^2_1+i\epsilon]}\frac{1}{[(k+q)^2-m^2_2+i\epsilon][q^2-m^2_3,+i\epsilon]}\label{Icaln}
\end{align}
with $n=0,1$. The integral in Eq.~(\ref{Icaln}) can be calculated analytically by using Cauchy's theorem, finding
\begin{align}
\mathcal{I}_n(m_1,m_2,m_3)=-i\frac{N_n(m_1,m_2,m_3)}{D(m_1,m_2,m_3)},\label{lcalnrat}
\end{align}
The $N_n$ and $D$ in Eq.~(\ref{lcalnrat}) depend on the energy of the particles involved in the loop and we refer the reader to Refs.~\cite{Malabarba:2020grf, Malabarba:2023zez} for more details. The integral in $d^3 q$ in Eq.~(\ref{QI}) can be obtained as
\begin{align}
\int\limits_{-\infty}^\infty \frac{d^3q}{(2\pi)^3}(\cdots)&\to\int\limits_0^\infty \frac{d|\vec{q}\,| |\vec{q}\,|^2}{(2\pi)^2}\int\limits_{-1}^{1}\text{dcos}\theta F(\Lambda,|\vec{k}+\vec{q}|)F(\bar \Lambda,|\vec{q}^{\,\text{CM}}|)(\cdots),\label{d3q}
\end{align}
where we consider $\vec{k}=|\vec{k}|\hat {z}$, and $\vec{q}=|\vec{q}\,|\text{sin}\theta(\text{cos}\phi\hat{i}+\text{sin}\phi\hat{j})+|\vec{q}\,|\text{cos}\theta\hat{k}$, such that $\vec{k}\cdot\vec{q}=|\vec{k}| |\vec{q}\,|\text{cos}\theta$ and, thus, the integral in $d\phi$ is trivial. In Eq.~(\ref{d3q}), $F$ represents a form factor introduced for the different vertices to take into account the finite size of $\phi(2170)$, $f_0(980)$, etc., and $\Lambda$, $\bar \Lambda$ are cutoffs of around 1000 MeV for the center-of-mass momentum of the particles forming these states. Typical expressions for the form factors in Eq.~(\ref{d3q}) are Lorentz~\cite{Gamermann:2009uq},
\begin{align}
F(\Lambda,|\vec{Q}|)=\frac{\Lambda^2}{\Lambda^2+|\vec{Q}|^2},
\end{align}
or Gaussian functions,
\begin{align}
F(\Lambda,|\vec{Q}|)=e^{-\frac{|\vec{Q}|^2}{2\Lambda^2}}.
\end{align}

Once  the coefficients appearing in the Lorentz expansion of the corresponding tensor integrals are determined, the partial decay width $\phi(2170)\to A B$ can be obtained by means of
\begin{align}
\Gamma_{\phi_R\to A B}=\frac{|\vec{p}_\text{CM}|}{24\pi m^2_{\phi_R}}\sum\limits_\text{pol}|t_{\phi_R\to AB}|^2,\label{wfor}
\end{align}
with $|\vec{p}_\text{CM}|$ being the modulus of the center-of-mass momentum of the particles in the final state and $\sum\limits_\text{pol}$ indicating the sum over the polarizations of the initial and final states.

\section{Results}
In Tables~\ref{TB1},~\ref{TB2}, and ~\ref{TB3} we show the results obtained within our description for the branching fractions
\begin{align}
B_1\equiv\frac{\Gamma_{\phi_R\to K^+(1460)K^-}}{\Gamma_{\phi_R\to K^+_1(1400)K^-}}=\frac{\mathcal{B}r[\phi_R\to K^+(1460)K^-]}{\mathcal{B}r[\phi_R\to K^+_1(1400)K^-]},\label{Br1}
\end{align}
\begin{align}
B_2\equiv\frac{\Gamma_{\phi_R\to K^+(1460) K^-}}{\Gamma_{\phi_R\to K^+_1(1270)K^-}}=\frac{\mathcal{B}r[\phi_R\to K^+(1460) K^-]}{\mathcal{B}r[\phi_R\to K^+_1(1270)K^-]},\label{Br2}
\end{align}
\begin{align}
B_3\equiv\frac{\Gamma_{\phi_R\to K^+_1(1270) K^-}}{\Gamma_{\phi_R\to K^+_1(1400)K^-}}=\frac{\mathcal{B}r[\phi_R\to K^+_1(1270) K^-]}{\mathcal{B}r[\phi_R\to K^+_1(1400)K^-]}.\label{Br3}
\end{align}
The values listed in the aforementioned tables can be compared with those obtained from the experimental values: in Ref.~\cite{BESIII:2020vtu}, the values (in eV) for the products $\mathcal{B}r\Gamma^{e^+e^-}_R$ are
\begin{align}
\mathcal{B}r[\phi_R\to K^+(1460)K^- ]\Gamma^{e^+e^-}_R&=3.0\pm 3.8,\nonumber\\
\mathcal{B}r[\phi_R\to K^+_1(1400)K^- ]\Gamma^{e^+e^-}_R&=\left\{\begin{array}{c}4.7\pm3.3,~\text{Solution 1}\\98.8\pm7.8,~\text{Solution 2}\end{array}\right.,\nonumber\\
\mathcal{B}r[\phi_R\to  K^+_1(1270)K^-]\Gamma^{e^+e^-}_R&=\left\{\begin{array}{c}7.6\pm3.7,~\text{Solution 1}\\152.6\pm14.2,~\text{Solution 2}\end{array}\right..\label{Brexp}
\end{align}
where two possible solutions for $\mathcal{B}r\Gamma^{e^+e^-}_R$ from the fits to the data were found in Ref.~\cite{BESIII:2020vtu} in case of the decays $\phi(2170)\to K^+_1(1400)K^- $, $K^+_1(1270)K^- $. Using Eq.~(\ref{Brexp}), we can obtain the experimental values for the $B_1$, $B_2$ and $B_3$ ratios of Eqs.~(\ref{Br1})-(\ref{Br3}), which are listed under the label ``Experiment'' in Tables~\ref{TB1}-\ref{TB3}. The theoretical values found for $B_1$, $B_2$, and $B_3$, as shown in Ref.~\cite{Malabarba:2020grf} do not depend much on the form factor considered in the vertices involved in the mechanisms depicted in Fig.~\ref{decay} and we provide here an average value of the results obtained with a Heaviside, a Lorentz and a Gaussian form factors. 

As can be seen in Tables~\ref{TB1}-\ref{TB3}, we find compatible results with the values extracted from the experiment, however, there is a strong dependence of these ratios on the particular model used to describe $K_1(1270)$ and $K_1(1400)$. More precise data would be required to distinguish whether $K_1(1270)$ is a state generated from the pseudoscalar-vector dynamics considered in Ref.~\cite{Geng:2006yb}. Note, however, that only if a superposition of the two poles obtained in Ref.~\cite{ Geng:2006yb} is considered in the calculation, a solution compatible with the value extracted from the experiment is obtained. Also, model B does not seem to give a good description of the ratio $B_2$. 
\begin{table}[h!]
\centering
\caption{Results for the branching ratio $B_1$.}\label{TB1}
\begin{tabular}{ccc}
\hline
&&$B_1$\\
\hline\hline
\multirow{2}{*}{Our results}&Model B&$0.62\pm0.20$\\
&Model C&$0.11\pm0.04$\\
\hline
\multirow{2}{*}{Experiment}&Solution 1&$0.64\pm0.92$\\
&Solution 2&$0.03\pm0.04$\\
\hline
\end{tabular}
\end{table}
\begin{table}[h!]
\centering
\caption{Results for the ratio $B_2$. In the case of model $C$, due to the uncertainty in the partial decay widths for $K_1(1270)$ and $K_1(1400)$ listed in Ref.~\cite{ParticleDataGroup:2022pth}, three different solutions were found in Ref.~\cite{Malabarba:2020grf} for the value of the coupling constant of $K_1$ to $\phi K$.}\label{TB2}
\begin{tabular}{cccl}
\hline
&&$B_2$\\
\hline\hline
\multirow{7}{*}{Our results}&\multirow{3}{*}{Model A}& $1.3\pm0.4$&(Poles $z_1$, $z_2)$\\
& &$3.6\pm1.2$&(Pole $z_1$)\\
& &$8.8\pm2.8$&(Pole $z_2$)\\
& Model B& $16\pm6$\\
& \multirow{3}{*}{Model C}& $1.2\pm0.4$&(Solution $\mathbb{S}_1$)\\
&  & $0.12\pm0.04$&(Solution $\mathbb{S}_2$)\\
& & $0.05\pm0.02$&(Solution $\mathbb{S}_3$)\\
\hline
\multirow{2}{*}{Experiment}&Solution 1&$0.40\pm0.54$&\\
&Solution 2&$0.02\pm0.03$&\\
\hline
\end{tabular}
\end{table}
\begin{table}[h!]
\centering
\caption{Results for the ratio $B_3$.}\label{TB3}
\begin{tabular}{cccl}
\hline
&&$B_3$\\
\hline\hline
\multirow{4}{*}{Our results}& Model B& $0.04\pm0.01$\\
& \multirow{3}{*}{Model C}& $0.09\pm0.02$&(Solution $\mathbb{S}_1$)\\
&  & $0.96\pm0.16$&(Solution $\mathbb{S}_2$)\\
& & $2.40\pm0.40$&(Solution $\mathbb{S}_3$)\\
\hline
\multirow{2}{*}{Experiment}&Solution 1&$1.62\pm1.38$&\\
&Solution 2&$1.55\pm0.19$&\\
\hline
\end{tabular}\end{table}

The results for the ratio $R_{\eta/\eta^\prime}$ between the widths of $\phi(2170)\to \phi\eta$ and to $\phi\eta^\prime$ are summarized in Table~\ref{R}. The results listed in this table should be compared with the ratio $R^\text{exp}_{\eta/\eta^\prime}\equiv \mathcal{B}^{\phi(2170)}_{\phi\eta}\Gamma^{\phi(2170)}_{e^+e^-}/\mathcal{B}^{\phi(2170)}_{\phi\eta^\prime}\Gamma^{\phi(2170)}_{e^+e^-}$ obtained by using the values $\mathcal{B}^{\phi(2170)}_{\phi\mathcal{P}}\Gamma^{\phi(2170)}_{e^+e^-}$ found in Refs.~\cite{BESIII:2021bjn,BESIII:2020gnc} (several solutions were found for this product by fitting the data):
\begin{align}
R^\text{exp}_{\eta/\eta^\prime}=\left\{\begin{array}{l}0.034^{+0.018}_{-0.011}~\text{solution I},\\ 1.42^{+0.58}_{-0.48}~\text{solution II,}\end{array}\right.\label{Rbes}
\end{align}
and with the results obtained for $R^\text{exp}_{\eta/\eta^\prime}$ by using for $\mathcal{B}^{\phi(2170)}_{\phi\eta}\Gamma^{\phi(2170)}_{e^+e^-}$ the value found in Ref.~\cite{Belle:2022fhh}, which gives
\begin{align}
R^\text{exp}_{\eta/\eta^\prime}=\left\{\begin{array}{l}0.013\pm0.007~\text{solution I},\\ 0.009\pm 0.003~\text{solution II},\\2.4\pm0.4~\text{solutions III, IV}.\end{array}\right.\label{RBB}
\end{align}

Considering the values listed in Table~\ref{R}, we find that mixing angles of $\simeq -22^\circ$ give rise to values for $R_{\eta/\eta^\prime}$ which are closer to the upper limit of the solution II of Eq.~(\ref{Rbes}) and solutions III, IV of Eq.~(\ref{RBB}).

\begin{table}[h!]
\centering
\caption{Values for the ratio $R_{\eta/\eta^\prime}$ considering different $\eta-\eta^\prime$ mixing angles, $\beta$, and form factors. The labels L and G indicate the consideration of a Lorentz (L) or a Gaussian (G) form factors, while the numbers I and II refer to the model used to calculate the $\mathcal{P}\bar{\mathcal{P}}^\prime$ $t$-matrix.}\label{R}
\begin{tabular}{ccccc}
$\beta$ (Degree)&&$-15$&$-19.47$&$-22$\\\\
\multirow{4}{*}{$R_{\eta/\eta^\prime}$}&LI&$5.12\pm 1.57$&$3.93\pm 1.21$ &$3.39\pm 1.04$ \\ 
&GI&$5.47\pm 1.68$&$4.21\pm 1.29$&$3.63\pm 1.11$\\
&LII&$4.21\pm1.29$&$3.25\pm1.00$&$2.80\pm0.86$\\
&GII&$4.41\pm1.35$&$3.40\pm1.04$&$2.93\pm0.90$
\end{tabular}
\end{table}

It is worth stressing that, despite the considerable experimental uncertainty obtained for the previously determined ratios, models considering $\phi(2170)$ as a $s\bar s$ states, a hybrid, etc., have real challenges in finding a good reproduction of these ratios, together with the mass and width of $\phi(2170)$.

\section{Conclusions}
In this work, we have summarized our findings for the branching ratios of $\phi(2170)$ to final states involving a $\bar K$ and a Kaonic resonance or a $\phi$ and an $\eta/\eta^\prime$ mesons. The description of $\phi(2170)$ as a $\phi f_0(980)$ molecular state produces values compatible with the experimental findings, reinforcing the interpretation of $\phi(2170)$ as a state generated by the three-body dynamics involved in the $\phi K\bar K$ system in isospin 0, with s-wave interactions and in which the $K\bar K$ subsystem resonates as $f_0(980)$. The values obtained for these ratios depend on the nature of the Kaonic resonances involved in the final state as well, and more precise data are needed to disentangle whether $K_1(1270)$ is a molecular state obtained from pseudoscalar-vector dynamics and the nature of $K_1(1400)$. 

\section{Acknowledgement}
This work is partly supported by the Brazilian agencies CNPq (Grant numbers 306461/2023-4, 304510/2023-8 and 407437/2023-1), and FAPESP (Grant Numbers 2020/00676-8, 2022/08347-9, 2023/01182-7).
%
% BibTeX or Biber users please use (the style is already called in the class, ensure that the "woc.bst" style is in your local directory)
 \bibliography{refs} % Replace "your_bib_file" with the actual name of your .bib file
%
% Non-BibTeX users please use
%
%\begin{thebibliography}{}
%
% and use \bibitem to create references.
%
%\bibitem{RefJ}
% Format for Journal Reference
%Journal Author, Article title. Journal \textbf{Volume}, page numbers (year). \url{https://doi.org/Article-DOI-number}
% Format for books
%\bibitem{RefB}
%Book Author, \textit{Book title} (Publisher, place, year) page numbers
% etc
%\end{thebibliography}

\end{document}